# When to Update the Sequential Patterns of Stream Data?

Qingguo Zheng  Ke Xu  Shilong Ma
National Lab of Software Development Environment
Department of Computer Science and Engineer
Beijing University of Aeronautics and Astronautics, Beijing 100083
{zqg, kexu, slam}@nlsde.buaa.edu.cn

## Abstract

In this paper, we first define a difference measure between the old and new sequential patterns of stream data, which is proved to be a distance. Then we propose an experimental method, called TPD (Tradeoff between Performance and Difference), to decide when to update the sequential patterns of stream data by making a tradeoff between the performance of increasingly updating algorithms and the difference of sequential patterns. The experiments for the incremental updating algorithm IUS on two data sets show that generally, as the size of incremental windows grows, the values of the speedup and the values of the difference will decrease and increase respectively. It is also shown experimentally that the incremental ratio determined by the TPD method does not monotonically increase or decrease but changes in a range between 20 and 30 percentage for the IUS algorithm.

## 1. Introduction

To enhance the performance of algorithms of data mining, many researchers [1,2,3,4,5,6,7] have focused on increasingly updating association rules and sequential patterns. But if we update association rules or sequential patterns too frequently, the cost of computation will increase significantly. For the problem above, Lee and Cheung [8] studied the problem "Maintenance of Discovered Association Rules: When to update?" and proposed an algorithm called DELL to deal with it. The important problem of "When to update" is to find a suitable distance measure between the old and new association rules. In [8], the symmetric difference was used to measure the difference of association rules. But Lee and Cheung only considered the difference of association rules, and did not consider that the performance of increasingly updating algorithms will change with the size of added transactions. Ganti et al. [15,16] focused on the incremental stream data mining model maintenance and change detection under block evolution. However, they also didn't consider the performance of incremental data mining algorithms for the evolving data.

Obviously, with the increment of the size of incremental windows, the performance of incremental updating algorithms will decrease. If the difference between the new and old sequential patterns is too high, the size of incremental window will become too large, therefore the performance of incremental updating algorithm will be reduced greatly. On the other hand, if the difference is too small, the incremental updating algorithm will update the old sequential patterns very frequently, which will also consume too many computing resource. In all, we must make a tradeoff between the performance of the updating algorithms and the difference of the sequential patterns. In this paper, we use the speedup as a measure of incremental updating algorithms and define a metric distance as the difference measure to detect the change of the sequential patterns of stream data. Based on those measures, we propose an experimental method, called TPD (Tradeoff between Performance and Difference), to estimate the suitable range of the incremental ratio of stream data by making a tradeoff between the performance of the increasingly updating algorithm and the difference of sequential patterns.

By the TPD method, we estimate the suitable range of incremental ratio of stream data for the incremental updating algorithm IUS [21]. The experiments on two data sets in Section 5 show that generally, as the size of incremental windows grows, the values of the speedup and the values of the difference will decrease and increase respectively. By the experiments, we can discover that as the size of original windows increases, the incremental ratio determined by TPD method does not monotonically increase or decrease but changes in a range between 18 and 30 percentage for the IUS algorithm.

The rest of this paper is organized as follows. In Section 2 we compare our work with the related work of the incremental mining for stream data. Section 3 formally gives the notion and definitions in this paper and defines the difference measure between the old and new sequential patterns. In Section 4, by making a tradeoff between the performance of increasingly updating algorithm and the difference of sequential patterns, we propose a method, called TPD (Tradeoff between Performance and Difference), to decide when to update sequential patterns of stream data. We experimentally verify the proposed method on two sets of GSM alarm data in Section 5. Finally, we give the conclusion in Section 6.

## 2. Related work

Lee and Cheung studied the problem of "When to update association rules" to avoid the overhead updating the rules too frequently in [8]. But they only studied updating association rules in the transaction data, while we focus on studying when to update the sequential patterns of stream data. They used the ratio $|L\Delta L`|/|L$ as the difference measure of association rules, but in this paper we not only define a difference measure of sequential patterns but also prove that this measure is a distance.

In [9], Agrawal et al. proposed an indexing method for time sequences for processing similarity queries, which uses the Discrete Fourier Transform (DFT) to map time sequences to the frequency domain. In [10], Agrawal proposed a new model of similarity of time sequences that captures the intuitive notion that two sequences should be considered similar if they have enough non-overlapping time-ordered pairs of subsequences those are similar.

In [11], Datar et al. studied the problem of maintaining aggregates and statistics over data streams, proposed the sliding window model and formulated the basic counting problem whose solution can be used as a building block for solving most of the problems.

Mannila et al. [13] used an edit distance notion for measuring the similarity of events sequences, which is defined as the cost of the cheapest possible sequence of operations that transforms a sequence to another. The edit distance can be computed using dynamic programming algorithm. Later Mannila and Seppanen [14] described a simple method for similarity search in sequences of events, which is based on the use of random projections. Das et al. [12] also introduced the notion of an external measure between attribute A and attribute B, defined by looking at the values of probe functions on sub-relations defined by A and B.

Ganti et al. [15] studied the incremental stream data mining model maintenance and change detection under block evolution. They adopted the FOCUS framework [16] for change detection, which measures the deviation between two datasets first by the class of decision tree models and then by the class of frequent itemsets.

Domingos and Hulten [17] introduced Hoeffding trees and proposed a method for learning online from the high-volume data streams that are increasingly common, called VFDT (very Fast Decision Tree learner). Later Hulten et al. [18] proposed an efficient algorithm for mining decision trees from continuously-changing data streams, based on the ultra-fast VFDT decision tree learner, which was called CVFDT. Corters and Preginbon [19] have discussed the statistical and computational aspects of deploying signature-based methods in the applications of telecommunications industry.

## 3. Problem Definition

### 3.1. Stream data model

**An stream tuple and its length, An stream queue and its length, An stream viewing window and its size**

1.) **An stream event** is defined as $E_i=<e_i, t_n>$, $i, n=1,2,3,\ldots$, where $e_i$ is an stream event type, including alarm event type, call details record type, financial market data type etc. and $t_n$ is the time of stream event type occurring.

2.) **An stream tuple** is defined as $Q_i=( ( e_{k1}, e_{k2}, \ldots\ldots , e_{km}), t_i )$, $i,m=1,2,3,\ldots$; $k1,k2,\ldots,km=1,2,3,\ldots$ , where '$e_{k1}, e_{k2},\ldots\ldots, e_{km}$' are the stream event types which concurrently occur at the time $t_i$ . An stream tuple can be represented by an stream event i.e. $Q_i=(< e_{k1}, t_i>,< e_{k2}, t_i> ,\ldots\ldots, <e_{km}, t_i>)=( E_{k1}, E_{k2}, \ldots\ldots , E_{km})$. If an stream tuple $Q_i$ only contains one stream type, then $Q_i=(( e_{k1}), t_i)=( < e_{k1}, t_i>)= E_{k1}$.

3.) **The length of stream tuple** is $|Q_i|=|( e_{k1}, e_{k2}, \ldots\ldots, e_{km})|=m$, which is the number of the stream event types contained in the stream tuple.

4.) **An stream queue** is defined as $S_{ij}=<Q_i, Q_{i+1},\ldots, Q_j>$ $i,j=1,2,3,\ldots$, where $t_i< t_{i+1} <\ldots<t_j$. An stream queue can be represented by an stream event i.e. $S_{ij}=<Q_i, Q_{i+1},\ldots,Q_j>=< (E_{i1},.., E_{ik}), (E_{i+1}),\ldots, (E_{j1}, \ldots, E_{jm})>= < (E_{i1},\ldots, E_{ik}), E_{i+1},\ldots, (E_{j1}, \ldots, E_{jm})>$, $i1,\ldots,ik ,j1,\ldots, jm=1,2,3,\ldots$ .

5.) **The length of stream queue** is defined as $| S_{ij}|=|<Q_i, Q_{i+1}, \ldots, Q_j>|=j-i+1$, which is equal to the number of stream tuple contained in stream queue.

6.) **An stream viewing window** is defined as $W_k$=<$Q_m$, …, $Q_n$|d=n-m+1>, where n≥m; k,m,n=1,2,3,⋯ .

7.) **The size of stream viewing window** is defined as $|W_k|$=|<$Q_m$,…, $Q_n$|d=n-m+1>|=d , where k,m,n=1,2,3,⋯ .

8.) Given an sequence $seq_m$ =< $e_{i1}$, $e_{i2}$,…,$e_{im}$ > and an stream viewing window $W_k$, **the times of the sequence $seq_m$** occurring in the $W_k$ are defined as

$$\text{occur}(seq_m, W_k) = |\text{ the times of } seq_m \text{ occurring in } W_k\ |$$

9.) Given an sequence $seq_m$ =< $e_{i1}$, $e_{i2}$,…,$e_{im}$ >, **the support of the sequence $seq_m$** in an stream viewing window $W_k$ is defined as

$$\text{support}(seq_m, W_k) = \frac{\text{occur}(seq_m, W_k)}{|W_k|}$$

Obviously, we have $\text{occur}(seq_m, W_k) = \text{support}(seq_m, W_k) \times |W_k|$

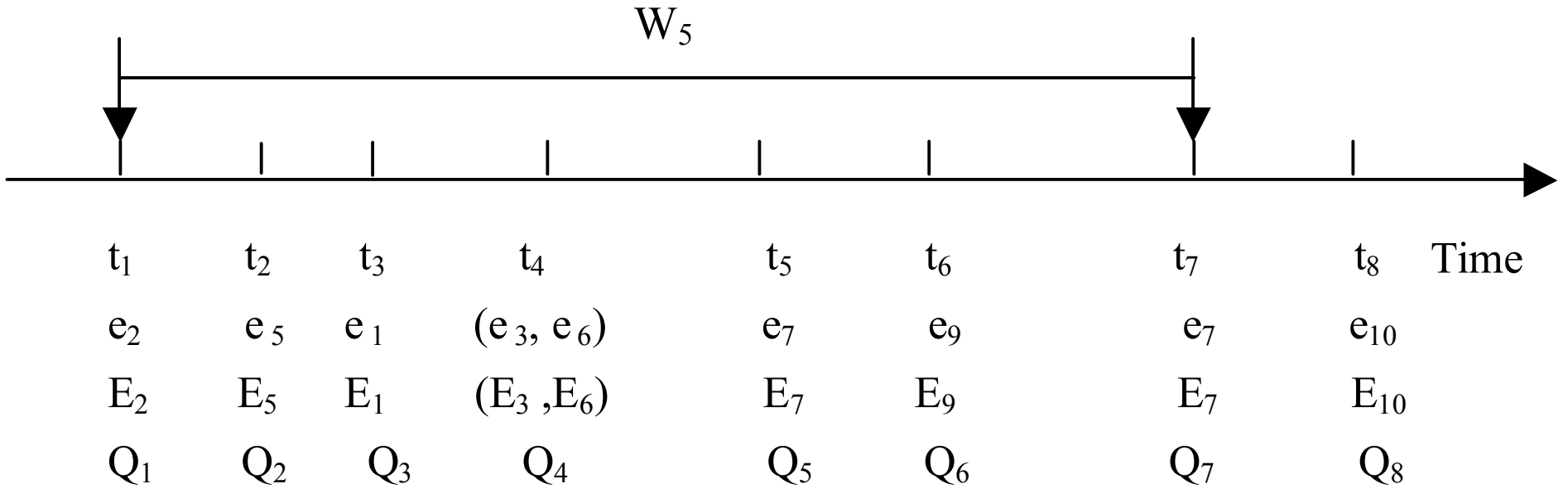

Figure 1. A stream queue

**Example 1** Figure 1 presents graphically an stream queue $S_{18}$=< $Q_1$, $Q_2$, $Q_3$, $Q_4$, $Q_5$, $Q_6$, $Q_7$, $Q_8$>. The time of stream tuple $Q_1$ , $Q_2$, … , $Q_8$ is that $t_1$< $t_2$ , … ,< $t_8$ .In Figure 1, the stream queue $S_{18}$ can also be denoted by the stream event, $S_{18}$=<$E_2$, $E_5$, $E_1$, ($E_3$ , $E_6$), $E_7$, $E_9$, $E_7$, $E_{10}$>. An stream viewing window is $W_5$=< $Q_1$, $Q_2$ , $Q_3$, $Q_4$, $Q_5$, $Q_6$, $Q_7$ | d=7> and the size of the window is $|W_5|$=7.

### 3.2. Sliding stream viewing windows on the stream queue

In Figure 2, Given stream view window $W_i$ (i=0,1,2,3, …), $\Delta W_i$ (i=0,1,2,3, …) is called **incremental window**, when i=0, $\Delta W_0$ is called **initial window**, where $W_{i+1}=W_i+\Delta W_{i+1}$ (i=0,1,2,3, …). The ratio of the size of incremental window to that of the initial window, i.e. $|\Delta W_1|/|W_0|$, is called **incremental ratio** of stream data.

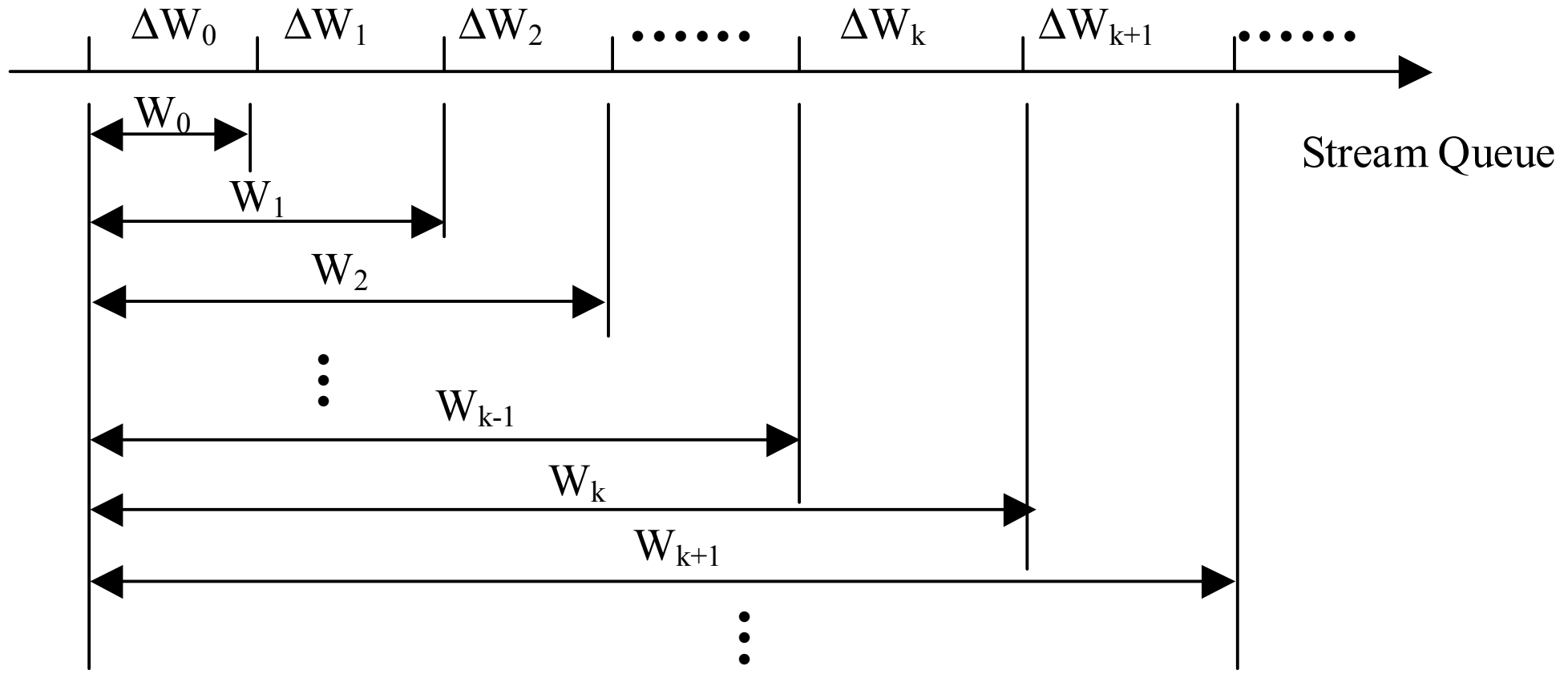

Figure 2. Sliding Windows

### 3.3. Estimation of the difference between the old and new sequential patterns

Before updating the frequent sequences $L^{W_K}$ in $W_k$, we must estimate the difference between $L^{W_K}$ and $L^{W_{K+1}}$. If the difference is very large, then we should update the sequential patterns as soon as possible. But if the difference is very small, then we do not need to update the sequential patterns. In order to measure the difference between the old and new frequent sequence sets, we can define a measure as follows

$$d(L^{W_K}, L^{W_{K+1}}) = \frac{|L^{W_K} \Delta L^{W_{K+1}}|}{|L^{W_K} \cup L^{W_{K+1}}|}, \text{ if } L^{W_K} \neq \Phi \text{ or } L^{W_{K+1}} \neq \Phi, \text{ otherwise } d(L^{W_K}, L^{W_{K+1}}) = 0,$$

where $L^{W_K} \Delta L^{W_{K+1}}$ is the symmetric difference between $L^{W_K}$ and $L^{W_{K+1}}$.

We know that a measure is not necessarily a metric. A metric distance must satisfy the triangle inequality which means that for three objects A, B and C in the metric space, the distance between A and C is greater than the distance between A and B plus the distance between B and C. This property is also useful in measuring the differences between frequent sequences because that for three frequent sequences $L^{w1}$, $L^{w2}$ and $L^{w3}$, if both the difference between $L^{w1}$ and $L^{w2}$ and the difference between $L^{w2}$ and $L^{w3}$ is very small, then, intuitively, the difference between $L^{w1}$ and $L^{w3}$ should also be small. If a difference measure between frequent sequences is a metric, then it follows from the triangle inequality that the above property holds. For the measure defined above, we can prove that it is also a metric (please see Appendix A).

## 4. The TPD (Tradeoff between Performance and Difference) method of deciding when to update sequential patterns

Lee and Cheung [8] only considered the difference between the old and new association rules, but that they didn't consider the change of the performance of incremental updating algorithms. As mentioned before, too large difference between the new and old sequential patters will result in poor performance of incremental updating algorithms, while too small difference will increase the computations lose significantly. Therefore, we must make a tradeoff between the difference of frequent sequences and the performance of increasingly updating algorithms and find the suitable range of the incremental ratio of stream data.

We propose an experimental method, called TPD (Tradeoff between Performance and Difference), to find the suitable range of incremental ratio of the initial window for deciding when to update sequential patterns of stream data. The TPD method uses the speedups as the measurement of incremental updating algorithms and adopts the measure defined in Section 3.3 as the difference between the new and old sequential patterns of stream data. With the increment of the size of incremental window, the speedup of the algorithm will decrease, while the difference of the old and new sequential patterns will increase. According to two main factors of the incremental updating algorithms, the TPD method maps the curve of the speedup and the difference changing with the size of incremental windows into the same graph under the same scale, and the points of intersection of the two curves are the suitable range of the incremental ratio of the initial windows for the increasingly updating algorithm.

In this paper, by the experiments in Section 5, we study the suitable range of incremental ratio of the initial windows for the incremental updating algorithm: IUS [21] by TPD (Tradeoff between Performance and Difference) method. In the experiments, the speedup ratio of the IUS algorithm is defined as **speedup=the execution time of Robust_search / the execution time of IUS**, where Robust_search is an algorithm to discover sequential patterns from stream data [20] and IUS is an increasingly updating sequential patterns algorithm based on Robust_search algorithm [21]. We use the distance i.e. $d(L^{W_k}, L^{W_{k+1}})$ defined above as the difference measure between the old frequent sequences $L^{W_k}$ and the new frequent sequences $L^{W_{k+1}}$.

The experiments of Section 5 show that generally, as the size of incremental windows grows, the values of the speedup and the values of the difference will decrease and increase respectively. By making data transform, called Min-max normalization [22], for the values of the speedup and the difference, we can map the speedup and the difference with the increment of the size of incremental windows into the same graph under the same scale, and then from the graph we can find the intersection point of two lines, obviously, by which we can compute the suitable range of incremental ratio of the initial window to update sequential patterns according to the differences and speedups mentioned above.

The Robust_search Algorithm does compute the support of sequences in the stream queue [20], especially, which can search the support of sequences from stream queue containing noise data. But because we mainly study the problem of when to update sequential patterns, we only consider the condition that the stream queue doesn't contain noise data in this paper.

The following is IUS (Incrementally Updating Sequences) algorithm proposed in [21]. In this paper, we rewrite the IUS algorithm using the stream data model. The input of the algorithm is the window $W_k$, , the incremental window $\Delta W_{k+1}$, and two parameters: Min_supp and Min_nbd_supp, where $W_{k+1}= W_k+\Delta W_{k+1}$. The output of the algorithm is the frequent sequences $L^{W_{K+1}}$ and the negative border sequences $NBD(W_{k+1})$.

**$W_k$**: The original stream view window which contains old time-related data.

**$\Delta W_{k+1}$**: The increment stream view window which contains new time-related data.

**$W_{k+1}$**: The updated stream view window. When stream data being increasingly updated, the total set of data which are equal to $W_k+\Delta W_{k+1}$

**Support(F, X):** the support of the sequence F in the X stream view windows, where $X \in \{ W_{k+1}, W_k, \Delta W_{k+1}\}$.

**Min_supp** :Minimum support threshold of the frequent sequence.

**Min_nbd_supp**: Minimum support threshold of negative border sequence.

**$C^X$**: Candidate sequences in X stream view windows, where $X \in \{ W_{k+1}, W_k, \Delta W_{k+1}\}$.

**$L^X$** : Frequent sequences in the X stream view windows, where $X \in \{ W_{k+1}, W_k, \Delta W_{k+1}\}$.

**NBD(X)**$=C^X- L^X$, where NBD(X) consists of the sequences in X stream view windows whose sub_sets are frequent, its Support is lower than Min_supp and greater than Min_nbd_supp. Note that $X \in \{W_{k+1}, W_k, \Delta W_{k+1}\}$

Figure 3. The Notions in IUS Algorithm

**Algorithm 1 . IUS(Incrementally Updating Sequences)**

Input: Frequent sequences set $L^{W_k}$, NBD($W_k$),$L^{\Delta W_{k+1}}$, NBD($\Delta W_{k+1}$), where $W_{k+1}= W_k+\Delta W_{k+1}$, Min_supp, Min_nbd_supp

Output: Frequent sequences set: $L^{W_{k+1}}$, Negative border sequences set: *NBD*($W_{k+1}$).

Generate $L_1^{W_{k+1}}$ from $L_1^{W_k}$, $L_1^{\Delta W_{k+1}}$, NBD($W_k$),NBD($\Delta W_{k+1}$);
m=2; L_Size=0;
while $((|L^{W_{k+1}}| - L_Size)>0)$
Begin
L_Size=$|L^{W_{k+1}}|$
For all $seq_m \in L^{W_k}$ and all subsets of $seq_m$ are frequent in $W_{k+1}$
{
If ($seq_m \in L^{\Delta W_{k+1}}$) Get occur($seq_m$ , $\Delta W_{k+1}$) from $L^{\Delta W_{k+1}}$ ;
Else If ($seq_m \in$ NBD($\Delta W_{k+1}$)) Get occur($seq_m$ , $\Delta W_{k+1}$) from NBD($\Delta W_{k+1}$);
Else search occur($seq_m$ , $\Delta W_{k+1}$) in $\Delta W_{k+1}$ ; /* compute frequent sequence in $L^{DB}$*/
If ((occur($seq_m$ , $\Delta W_{k+1}$)+ occur($seq_m$ , $W_k$))>(Min_supp• | $W_{k+1}$|))
Insert $seq_m$ into $L^{W_{k+1}}$;
else { Prune the $seq_m$ and the sequence containing $seq_m$ from $L^{W_k}$ and NBD($W_k$);
If((occur($seq_m$ , $\Delta W_{k+1}$)+ occur($seq_m$ , $W_k$))>(Min_nbd_supp• |$W_{k+1}$|))
Insert $seq_m$ into NBD($W_{k+1}$);
}
}
For all $seq_m \in L^{\Delta W_{k+1}}$ and $seq_m \notin L^{W_k}$ and all subsets of $seq_m$ are frequent in $W_{k+1}$

```
{
 If(seq_m∈NBD(W_k)) Get occur(seq_m , W_k) from NBD(W_k);
      Else search  occur(seq_m , W_k)  in W_k ;/* compute frequent sequence in L^{ΔW_{k+1}} */
 If( (occur(seq_m , ΔW_{k+1})+ occur(seq_m , W_k))>(Min_supp• |W_{k+1}|) )
                          Insert  seq_m  into  L^{W_{k+1}} ;
      else { Prune the seq_m and the sequence containing seq_m from L^{ΔW_{k+1}} and NBD(ΔW_{k+1});
              If((occur(seq_m , ΔW_{k+1})+ occur(seq_m , W_k))>(Min_nbd_supp• | W_{k+1}|))
                                     Insert seq_m into NBD(W_{k+1});
             }
}
For all seq_m∈NBD(W_k) and seq_m ∉L^{ΔW_{k+1}} and seq_m ∉NBD(ΔW_{k+1})
    and  all subset of seq_m are frequent
{   Search  occur(seq_m , ΔW_{k+1})  in  ΔW_{k+1};
     If((occur(seq_m , ΔW_{k+1})+ occur(seq_m , W_k))>(Min_nbd_supp• | W_{k+1} |))
                              insert seq_m  into  NBD(W_{k+1});
}
For all seq_m∈NBD(ΔW_{k+1}) and seq_m∉L^{W_k} and seq_m∉NBD(W_k)
   and all subset of seq_m are frequent
  {    Search  occur(seq_m , W_k)  in  W_k
        If((occur(seq_m , ΔW_{k+1})+ occur(seq_m , W_k))>(Min_nbd_supp•|W_{k+1}|))
                              insert  seq_m  into  NBD(W_{k+1});
  }
 / * generate new negative border from L^{ΔW_{k+1}} and L^{W_k}  */
  generate Negative Border NBD_m( W_{k+1} ). from L^{W_{k+1}}_{m-1};   /* Algorithm 1_1 */
m=m+1;
end.  /* end of while */
```

## 5. Experiments

We conducted a set of experiments to find when to update sequential patterns for stream data. The experiments were on the DELL PC Server with 2 CPU Pentium II, CPU MHz 397.952211, Memory 512M, SCSI Disk 16G. The Operating system on the server is Red Hat Linux version 6.0.

The data_1 in experiments are the alarms in GSM Networks, which contain 194 alarm types and 100k alarm events. The time of alarm events in the data_1 range from 2001-08-11-18 to 2001-08-13-17. The data_2 in experiments are the alarms in GSM Networks, which contain 171 alarm types and 100k alarm events. The time of alarm events in the data_2 range from 2001-08-07-09 to 2001-08-09-12.

In the experiments, we compare the execution time of the IUS algorithm with that of the Robust_search algorithm [21] in the view window $W_i$. The speedup ratio is defined in Section 4. We also use the distance in Section 3.3 as the difference measure between the sequential patterns of $W_k$ and $W_{k+1}$.

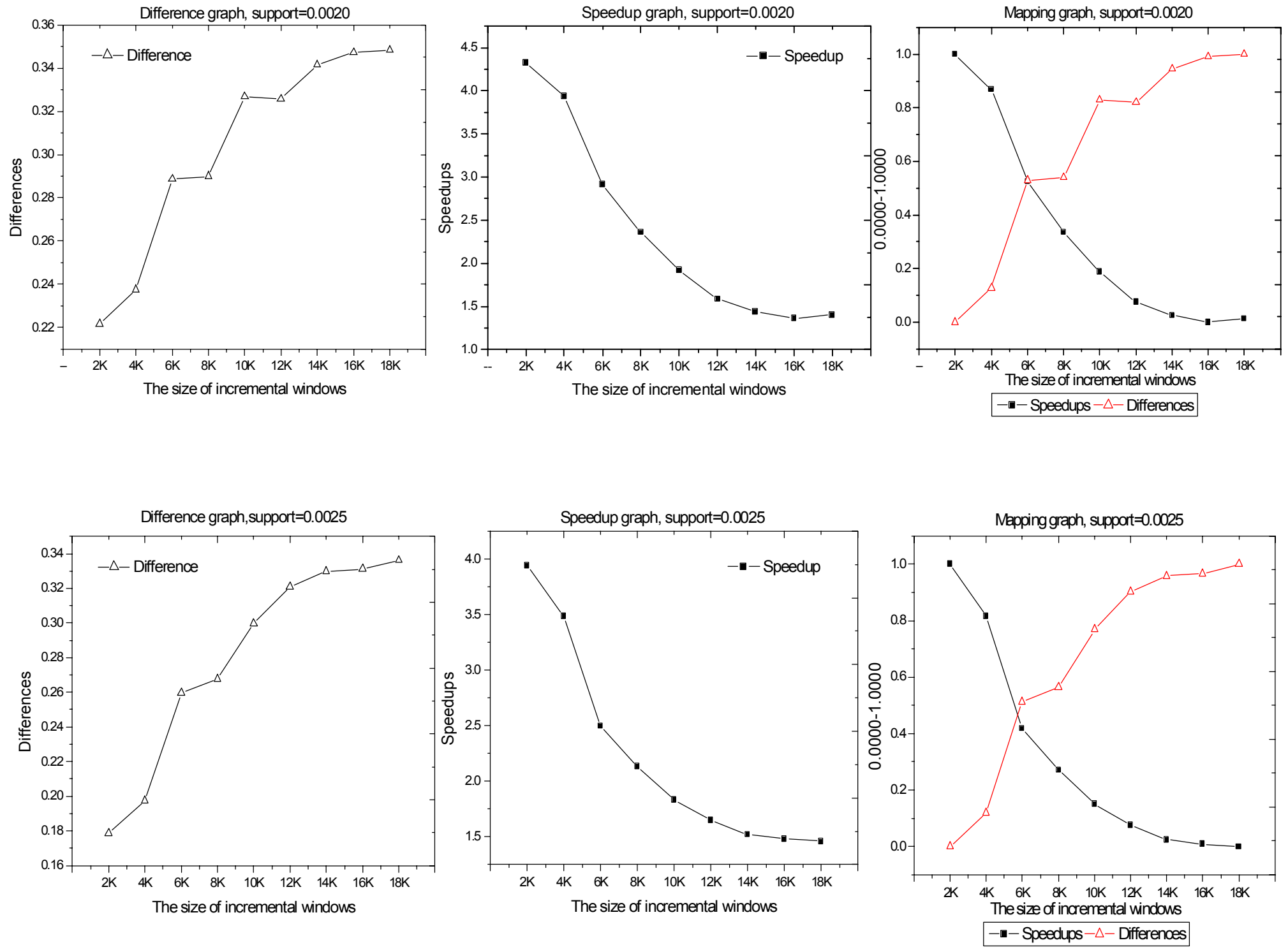

Figure 4.Experiment 1 on Data_1 |Initial window|=20k

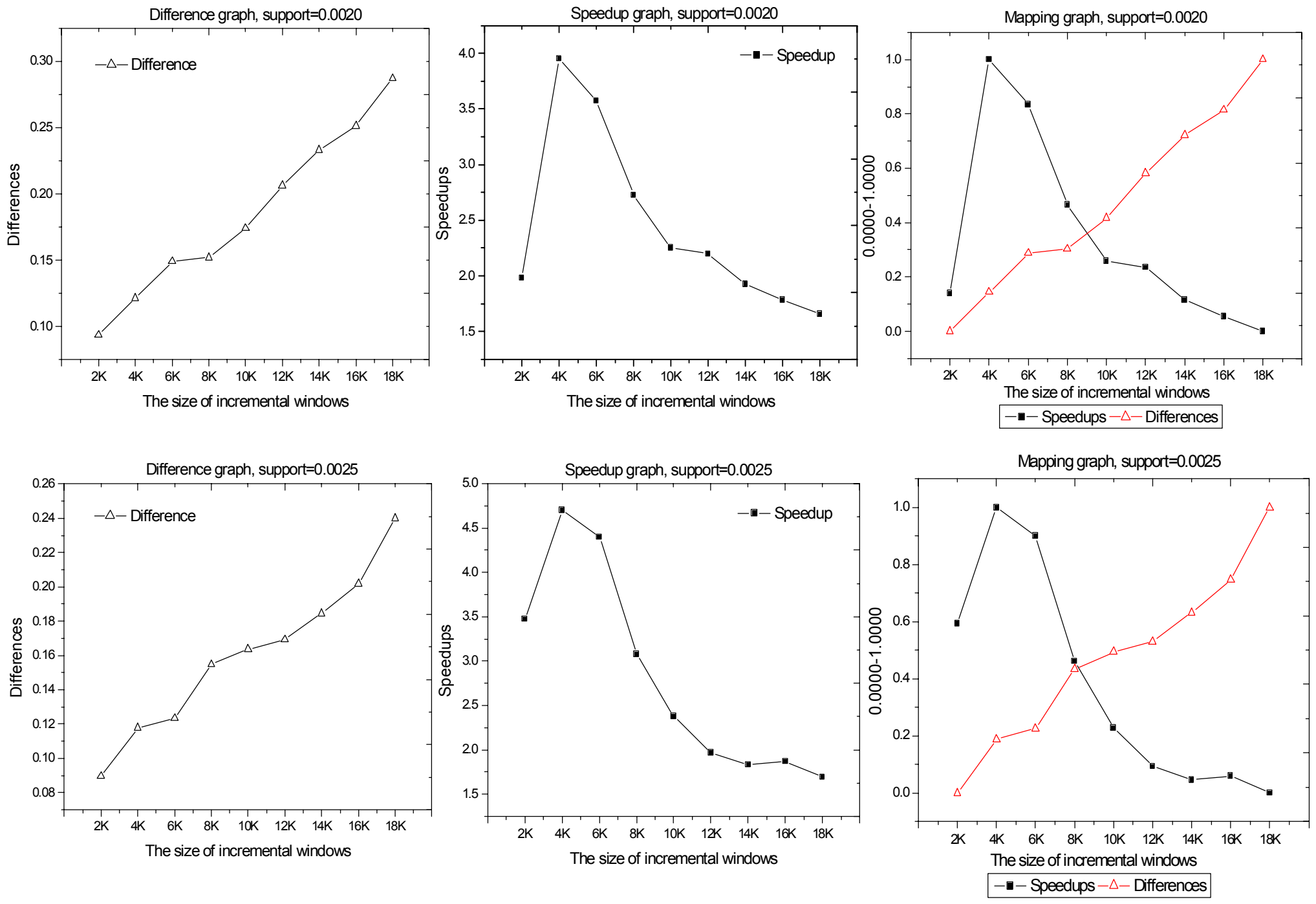

Figure 5. Experiment 2 on Data_2 |Initial window|=20K

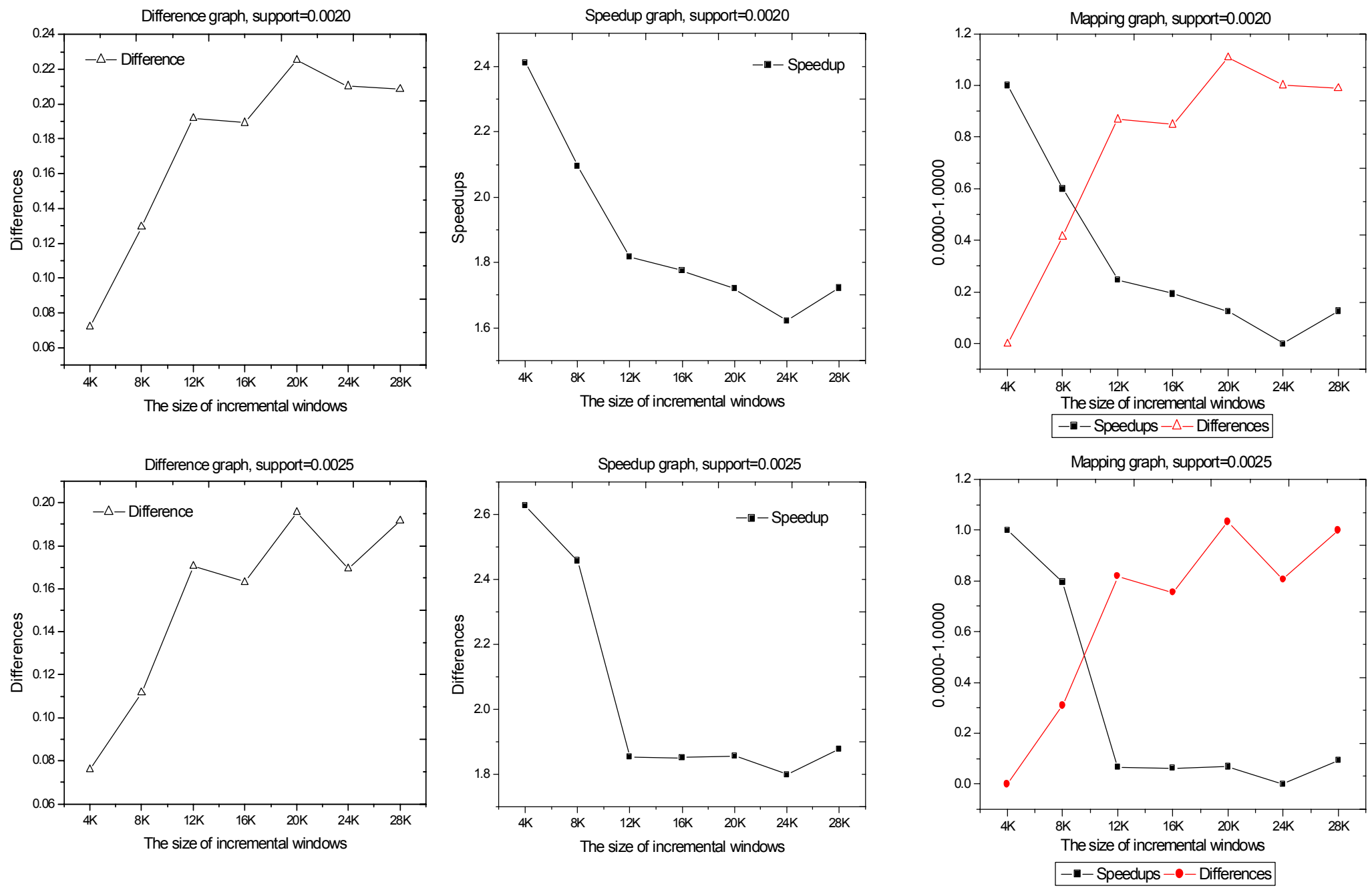

Figure 6. Experiment 3 on Data_1 |Initial window|=40K

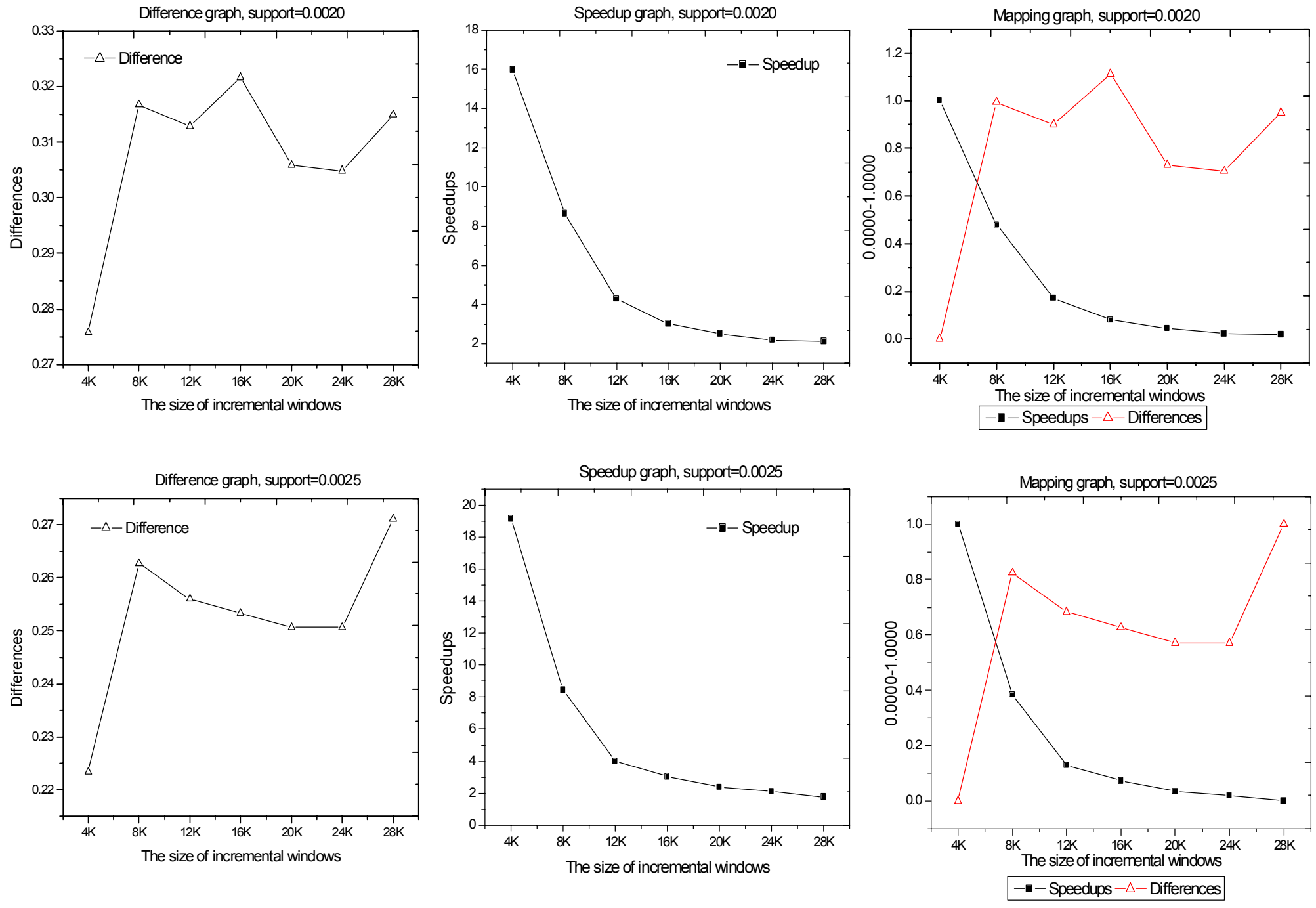

Figure 7. Experiment 4 on Data_2 |Initial window|=40K

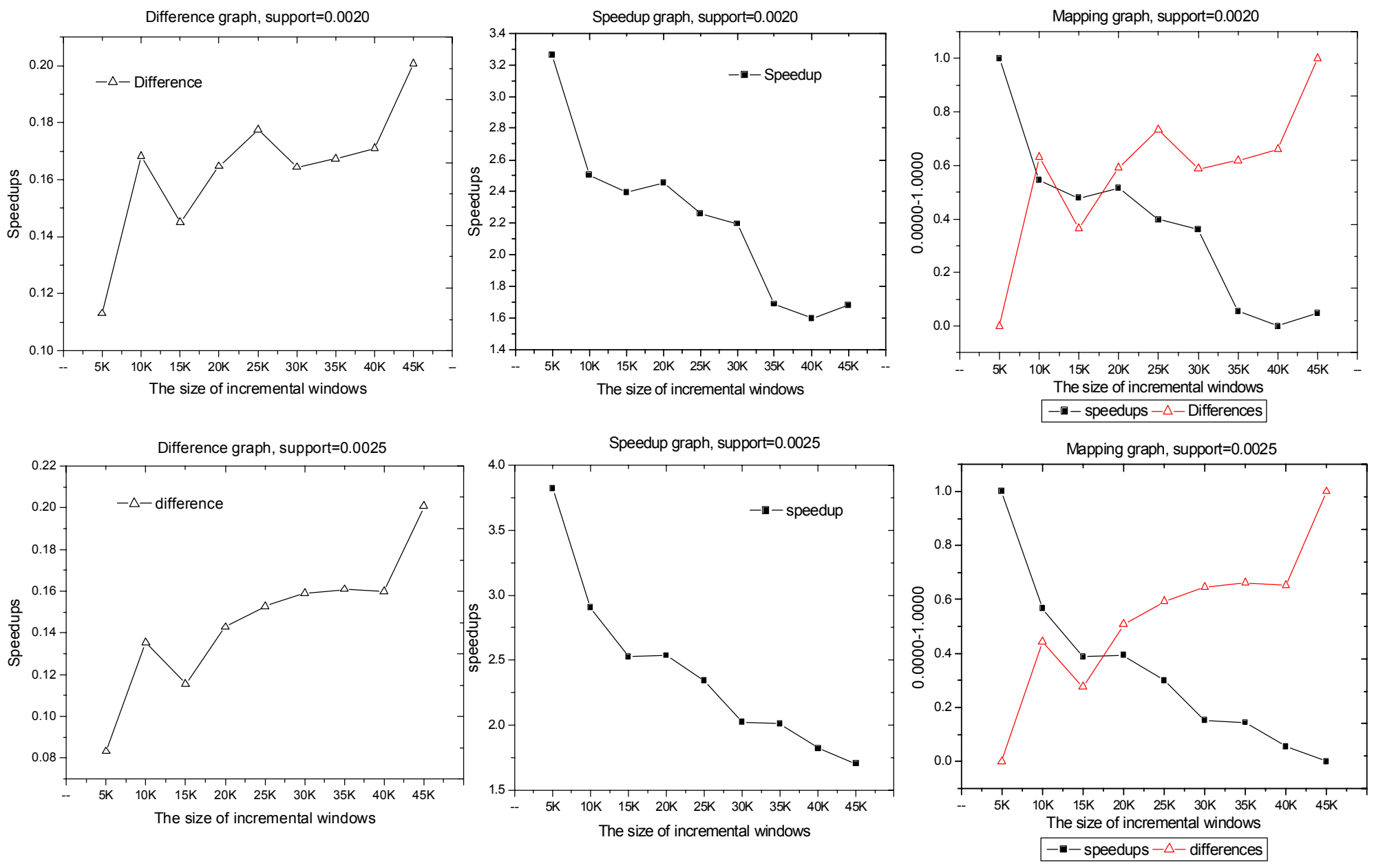

Figure 8. Experiment 5 on data_1 |Initial

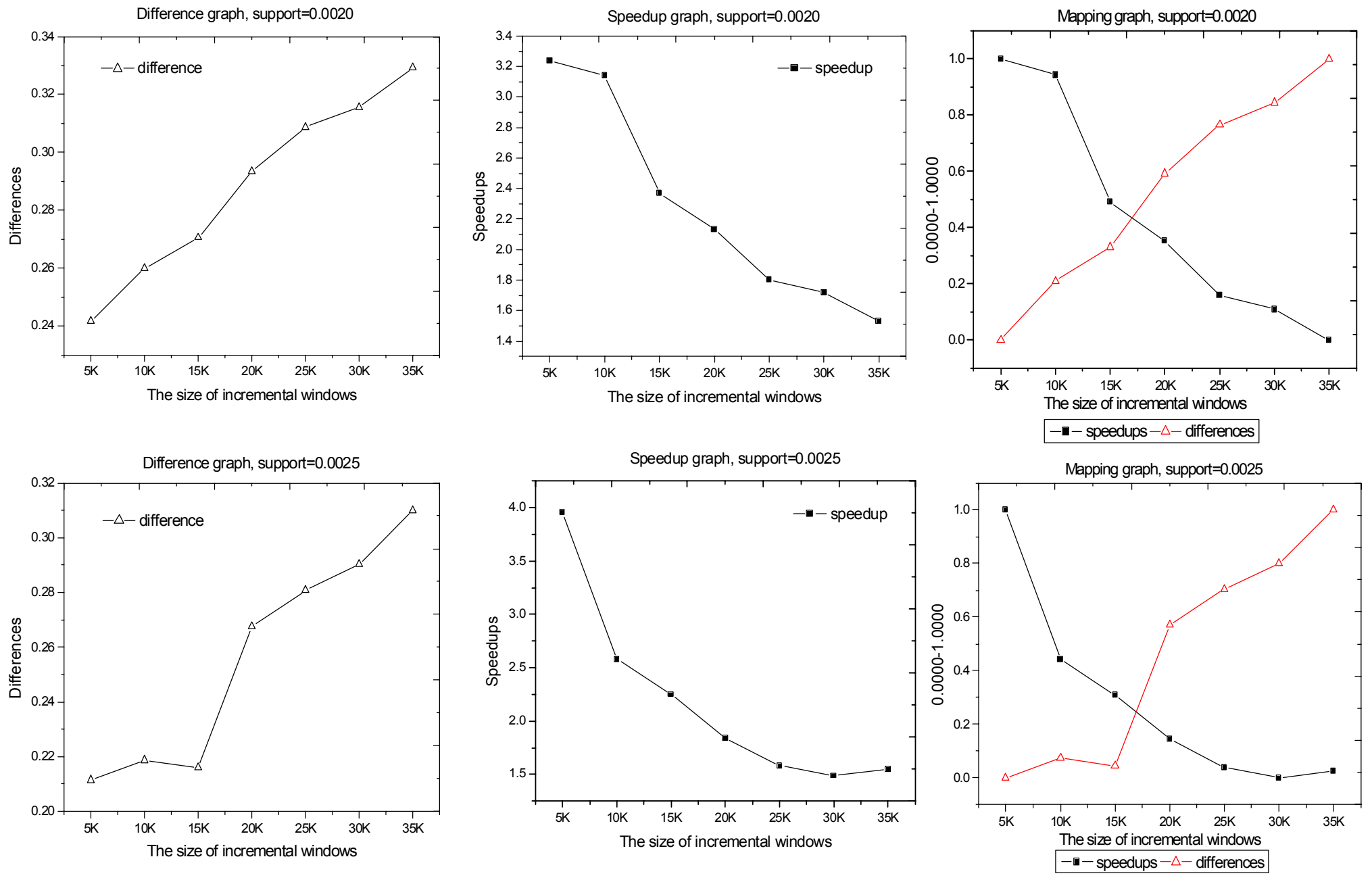

Figure 9. Experiment 6 on data_2 |Initial window|=50K

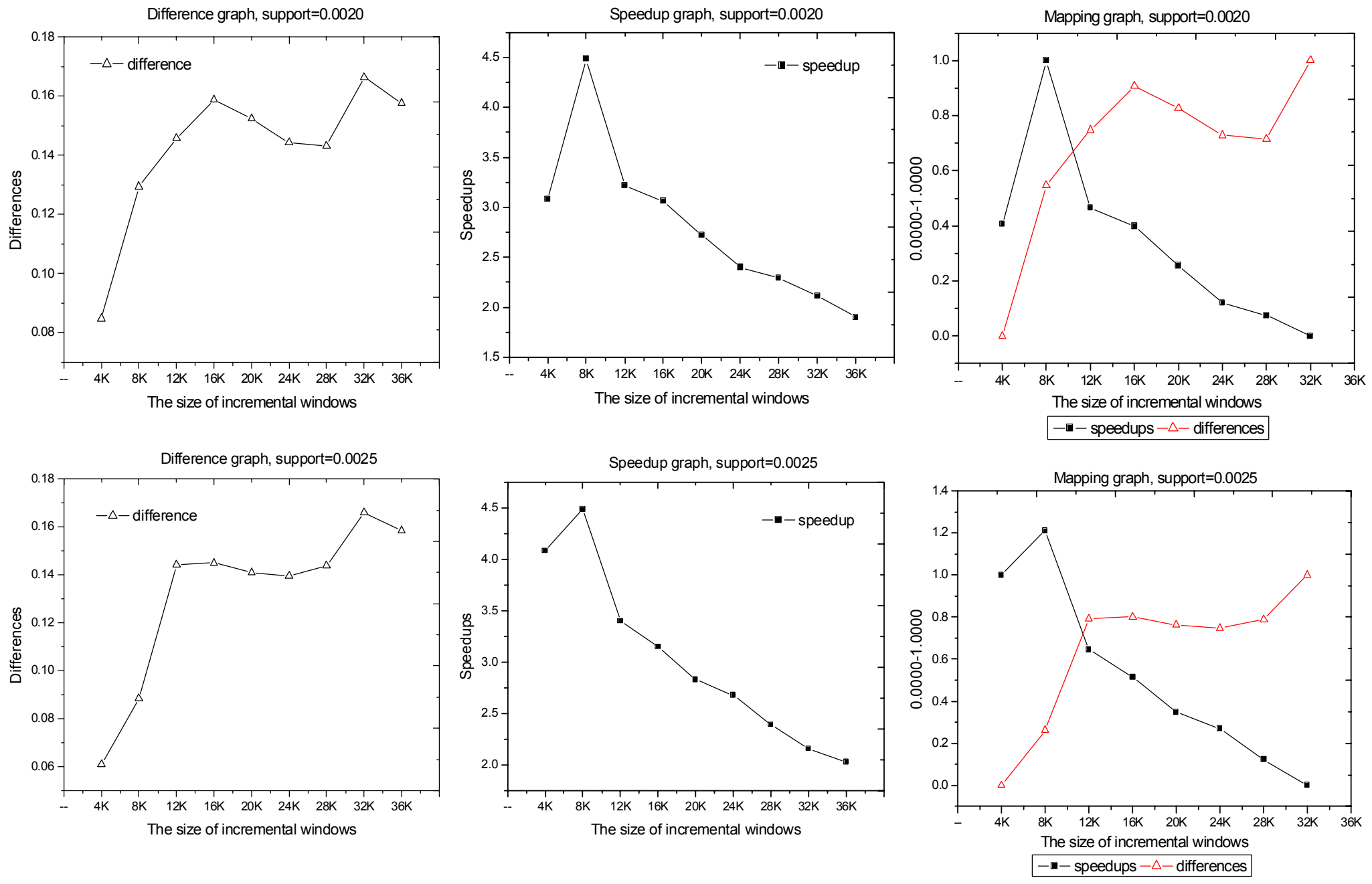

Figure 10. Experiment 7 on Data_1 |Initial window|=60K

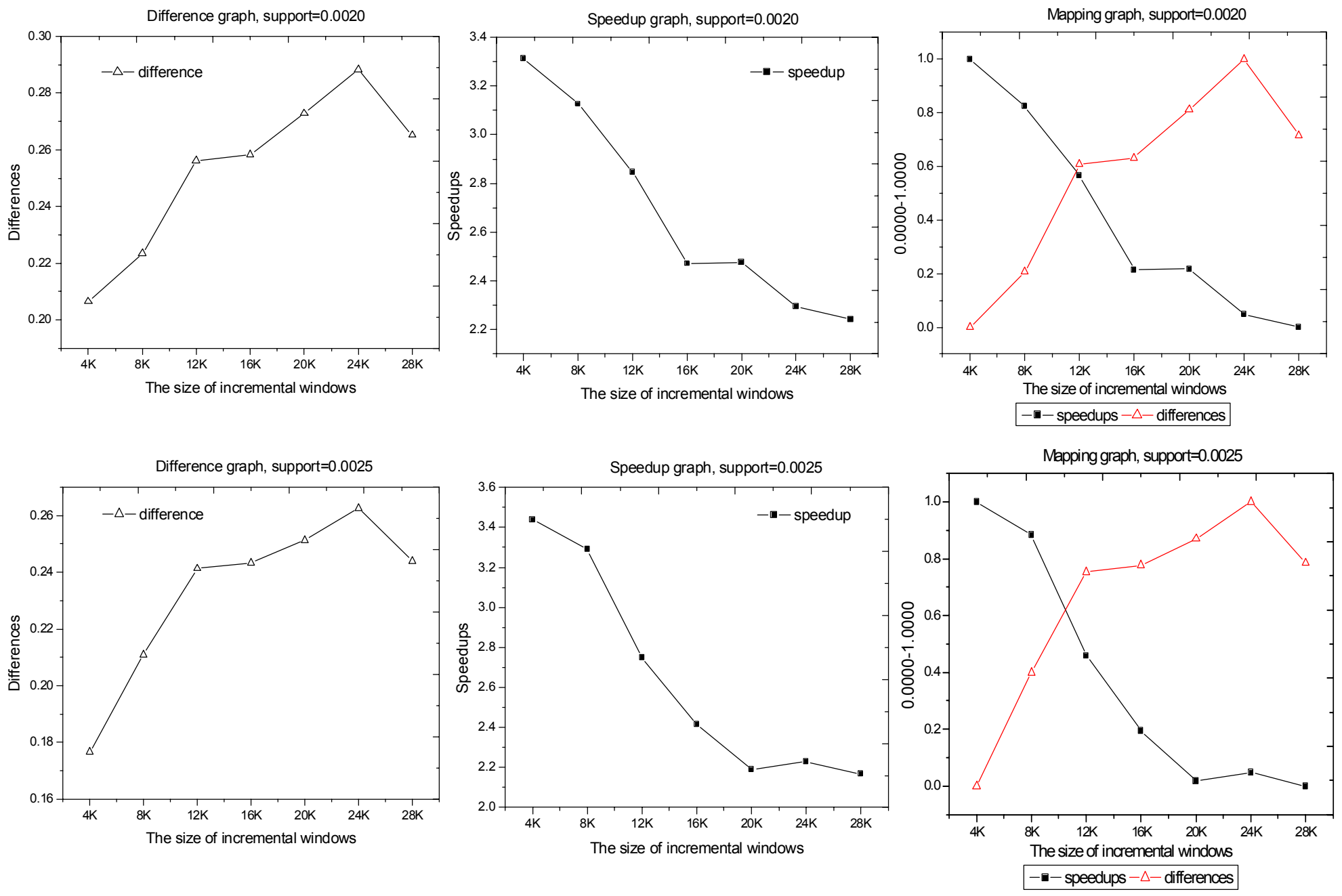

Figure 11. Experiment 8 on Data_2 |Initial window|=60K

The Speedup graph in Figure 4, 5, ⋯, 11 is the speedup of IUS Algorithm [20] to the Robust_search Algorithm [21] with the size of incremental windows. The Difference graph in Figure 4, 5 , ⋯, 11 is the difference measure of frequent sequences between the initial window and the incremental windows with the size of incremental windows. Mapping the Speed and difference graphs into the same scale graph forms Mapping graph in Figure 4, 5, ⋯ , 11.

In order to make the speed graph and difference graph have the same scale, we adopt the data normalization methods called Min-max normalization [22], which performs linear transformation of the origin data. Suppose that $min_A$ and $max_A$ are the minimum and maximum values of an attribute A. Min-max normalization maps a value ν of A to ν` in the range [$new_min_A$, $new_max_A$] by computing

$$v` = \frac{v - \min_A}{\max_A - \min_A}\left(new_\max_A - new_\min_A\right) + new_\min_A$$

In the experiments of this paper, we map the value of difference and speedup into the same range [0,1] by let $new_min_A$=0 and $new_max_A$=1. By data transformation above, we could map the broken lines of the difference and speedup into the same graph i.e. Mapping Graph under the same scale. The intersection of the two broke lines is the tradeoff size of the incremental window between the difference and speedup, by which we can compute the proper incremental ratio of incremental windows.

In the experiment 1 on data_1, we choose the initial window $|W_0|$=20K, and update the initial sequential patterns by the incremental size of 2K, 4K, 6K, 8K, 10K, 12K, 14K, 16K, and 18K, i.e. the size of incremental window $\Delta W_i$. The results of experiment 1 are illustrated in Figure 4. In the speedup graphs with support=0.0020 and with support=0.0025, the values of the speedup of IUS algorithm will decrease with the increment of size of incremental windows $\Delta W_i$ . In the difference graphs with support=0.0020 and with support=0.0025, the values of the difference will increase with the increment of the size of incremental alarm windows. In order to find the suitable size of incremental windows, we first map the graphs of speedup and difference into the same graph by the data transform above, then find the intersection point of the two lines. The intersection point is a tradeoff between the speedup and the difference, and is a suitable point to update sequential patterns. In the mapping graphs with support=0.0020 and support=0.0025 of Figure 4, the intersection point is about 6K, so the suitable range of incremental ratio of initial window is about 30 percent of initial windows $W_0$.

In the experiment 2 on data_2, we choose the initial window $|W_0|$=20K, and update the initial sequential patterns by the incremental size of 2K, 4K, 6K, 8K, 10K, 12K, 14K, 16K, and 18K, i.e. the size of incremental window $\Delta W_i$. The results of experiment 2 are illustrated in Figure 5. In the mapping graphs with support=0.0020 and support=0.0025 of Figure 5, the intersection point is about between 8.5K and 9K, so the suitable range of incremental ratio of initial window is about 42.5 to 45 percent of initial windows $W_0$

In the experiment 3 on data_1, we choose the initial window $|W_0|$=40K, and update the initial sequential patterns by the incremental size of 4K, 8K, 16K, 20K, 24K, 28K, and 32K, i.e. the size of incremental window $\Delta W_i$. The results of experiment 3 are illustrated in Figure 6. The intersection point is a tradeoff between the speedup and the difference, and is a suitable point to update sequential patterns. In the mapping graphs with support=0.0020 and support=0.0025 of Figure 6, the intersection point is between 9K and 10K, so the suitable range of incremental ratio of initial window is about 22.5 to 25 percent of initial windows $W_0$.

In the experiment 4 on data_2, we choose the initial window $|W_0|$=40K, and update the initial sequential patterns by the incremental size of 4K, 8K, 16K, 20K, 24K, 28K, and 32K, i.e. the size of incremental window $\Delta W_i$. The results of experiment 4 are illustrated in Figure 7. In the mapping graphs with support=0.0020 and support=0.0025 of Figure 7, the intersection point is about 6K, so the suitable range of incremental window of initial window is about 15 percent of initial windows $W_0$

In the experiment 5 on data_1, we choose the initial window $|W_0|$=50K, and update the initial patterns by incremental size of 5K, 10K, 15K, 20K, 25K, 30K, 35K 40k, and 45k, i.e. the size of incremental window $\Delta W_i$. The results of experiment 5 are illustrated in Figure 8. In the mapping graphs with support=0.0020 and support=0.0025 of Figure 8, the intersection point is between 15K and 18K, so the suitable range of incremental window of initial window is about 30 to 36 percent of initial windows $W_0$.

In the experiment 6 on data_2, we choose the initial window $|W_0|$=50K, and update the initial sequential patterns by the incremental size of 5K, 10K, 15K, 20K, 25K, 30K, and 35K, i.e. the size of incremental window $\Delta W_i$. The results of experiment 6 are illustrated in Figure 9. In the mapping graphs with support=0.0020 and support=0.0025 of Figure 9, the intersection point is between 15K and 18K, so the range of incremental window of initial window is about 30 to 36 percent of initial windows $W_0$.

In the experiment 7 on data_1, we choose the initial window $|W_0|$=60K, and update the initial sequential patterns by the incremental size of 4K, 8K, 16K, 20K, 24K, 28K, 32K, and 36K, i.e. the size of incremental window $\Delta W_i$. The results of experiment 7 are illustrated in Figure 10. In the mapping graphs with support=0.0020 and support=0.0025 of Figure 10, the intersection point is between 10K and 12K, so the suitable range of incremental window of initial window is about 16.7 to 20 percent of initial windows $W_0$

In the experiment 8 on data_2, we choose the initial window $|W_0|$=60K, and update the initial sequential patterns by the incremental size of 4K, 8K, 16K, 20K, 24K, 28K, 32K, and 36K, i.e. the size of incremental window $\Delta W_i$. The results of experiment 8 are illustrated in Figure 11. In the mapping graphs with support=0.0020 and support=0.0025 of Figure 11, the intersection point is between 10K and 12K, so the suitable range of incremental ratio of initial window is about 16.7 to 20 percent of initial windows $W_0$.

In all, by the experiments above, in general, as the size of incremental windows grows, the values of the speedup and the values of the difference will decrease and increase respectively. Based on the TPD method we proposed, it is shown experimentally that the suitable range of incremental ratio of initial windows to update is about 20 to 30 percent of the size of initial windows for the IUS algorithm.

## 6. Conclusion

In this paper, we first proposed a metric distance as the difference measure between the sequential patterns. Then we present an experimental method, called TPD (Tradeoff between Performance and Difference), to decide when to update sequential patterns of stream data. The TPD method can determine a reasonable ratio of the size of incremental window to that of original window for incremental updating algorithms, which may depend on the concrete data. We also do eight experiments of IUS algorithm [20] to verify the TPD method. From the experiments, we can see that as the size of original windows increases, the incremental ratio determined by the TPD method does not monotonically increase or decrease but changes in a range between 20 and 30 percentage.

So in practice, when we do incremental data mining for some kind of stream data, by use of the TPD method, we can do some initial experiments to determine a suitable incremental ratio for this kind of data and then use this ratio to decide when to update sequential patterns in the incremental data mining. Finally, we hope that our method could be extended to some other increasingly updating algorithms in future.

**ACKNOWLEDGEMENTS**

Thanks Professor Jiawei Han's Summer Course in Beijing 2002, help us to find the proper data transform methods. Thanks Professor Wei Li for the choice of subject and guidance of methodology. Thanks for the suggestions from Professor YueFei Sui of Chinese Academy Sciences. The author would like to thank other members of National Lab of Software Development Environment. This research was supported by **National 973 Project of China Grant No.G1999032701 and No.G1999032709**.

## Appendix A

Theorem 1. Given two sets $A$ and $B$, the measure defined as follows is a distance.

$$d(A,B)=\frac{|A\Delta B|}{|A\cup B|} \quad \text{if} \quad A\neq\Phi \quad \text{or} \quad B\neq\Phi\text{, otherwise} \quad d(A,B)=0\,.$$

*Proof.* First, we can easily prove that the following two properties hold for the measure defined above.

1. $d(A,B)\geq 0$ and $d(A,B)=0 \Leftrightarrow A=B$.
2. $d(A,B)=d(B,A)$.

Now we only need to prove that the triangle inequality also holds for this measure. Namely, given three sets $A$, $B$ and $C$, we will prove that

$$d(A,C)+d(B,C)\geq d(A,B)\,. \tag{A.1}$$

Assume that the sets $A$, $B$ and $C$ are as follows.

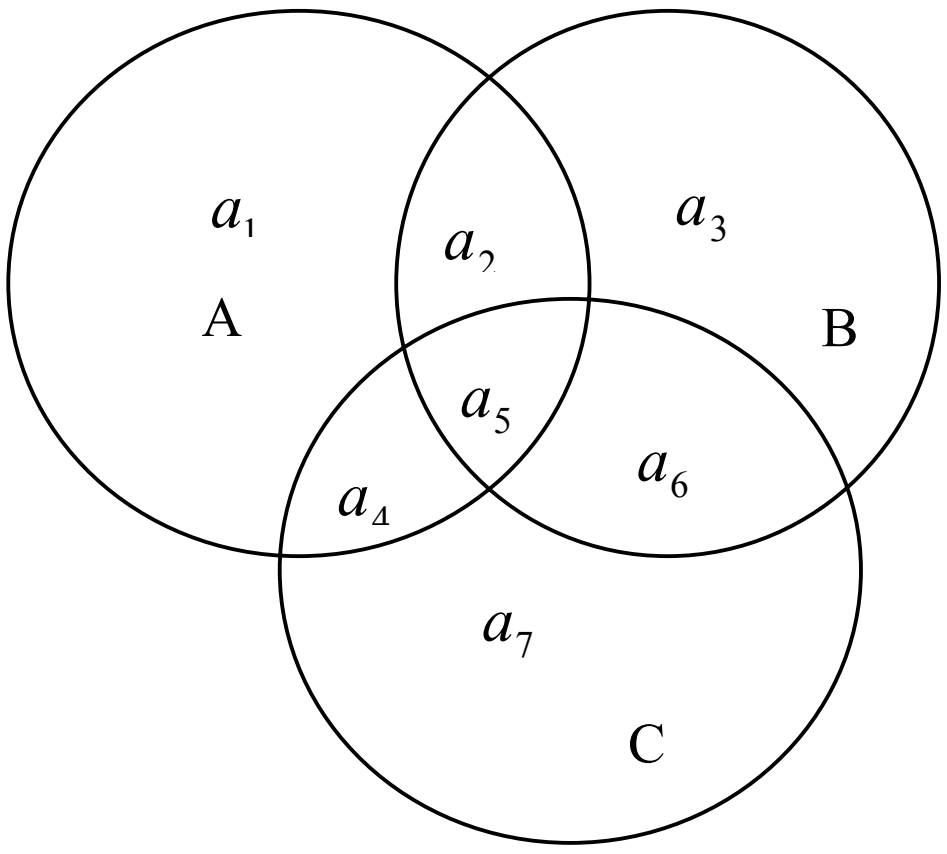

By the figure above and the definition of the measure $d$, we know that proving inequality (A.1) is equivalent to proving the following inequality.

$$\frac{a_1+a_2+a_6+a_7}{a_1+a_2+a_4+a_5+a_6+a_7}+\frac{a_2+a_3+a_4+a_7}{a_2+a_3+a_4+a_5+a_6+a_7}$$
$$\geq\frac{a_1+a_3+a_4+a_6}{a_1+a_2+a_3+a_4+a_5+a_6}\,. \tag{A.2}$$

It can be easily shown that if $b_1\geq b_3\geq 0$, $b_2\geq b_4\geq 0$ and $b_1+b_4>0$, then

$$\frac{b_3+b_2}{b_1+b_2}\geq\frac{b_3+b_4}{b_1+b_4}\,. \tag{A.3}$$

The proof of inequality (A.2) can be divided into the following four cases.

Case 1: $a_7\leq a_3$ and $a_7\leq a_1$.

By the condition above, we have

$$\frac{a_1+a_2+a_6+a_7}{a_1+a_2+a_4+a_5+a_6+a_7}\geq\frac{a_1+a_2+a_6+a_7}{a_1+a_2+a_4+a_5+a_6+a_3} \tag{A.4}$$

$$\frac{a_2+a_3+a_4+a_7}{a_2+a_3+a_4+a_5+a_6+a_7} \geq \frac{a_2+a_3+a_4+a_7}{a_2+a_3+a_4+a_5+a_6+a_1} \tag{A.5}$$

It follows from inequalities (A.4) and (A.5) that inequality (A.2) holds.

Case 2: $a_7 \geq a_3$ and $a_7 \geq a_1$.

By the condition above and inequality (A.3), we have

$$\frac{a_1+a_2+a_6+a_7}{a_1+a_2+a_4+a_5+a_6+a_7} \geq \frac{a_1+a_2+a_6+a_3}{a_1+a_2+a_4+a_5+a_6+a_3} \tag{A.6}$$

$$\frac{a_2+a_3+a_4+a_7}{a_2+a_3+a_4+a_5+a_6+a_7} \geq \frac{a_2+a_3+a_4+a_1}{a_2+a_3+a_4+a_5+a_6+a_1} \tag{A.7}$$

It follows from inequalities (A.6) and (A.7) that inequality (A.2) holds.

Case 3: $a_3 \leq a_7 \leq a_1$.

Since $a_3 \leq a_7$, by inequality (A.3) we have

$$\frac{a_1+a_2+a_6+a_7}{a_1+a_2+a_4+a_5+a_6+a_7} \geq \frac{a_1+a_2+a_6+a_3}{a_1+a_2+a_4+a_5+a_6+a_3} \tag{A.8}$$

$$\frac{a_2+a_3+a_4+a_7}{a_2+a_3+a_4+a_5+a_6+a_7} \geq \frac{a_2+a_3+a_4+a_3}{a_2+a_3+a_4+a_5+a_6+a_3} \tag{A.9}$$

Since $a_3 \leq a_1$, we have

$$\frac{a_2+a_3+a_4+a_3}{a_2+a_3+a_4+a_5+a_6+a_3} \geq \frac{a_2+a_3+a_4+a_3}{a_2+a_3+a_4+a_5+a_6+a_1} \tag{A.10}$$

By inequalities (A.9) and (A.10), we have

$$\frac{a_2+a_3+a_4+a_7}{a_2+a_3+a_4+a_5+a_6+a_7} \geq \frac{a_2+a_3+a_4+a_3}{a_2+a_3+a_4+a_5+a_6+a_1} \tag{A.11}$$

It follows from inequalities (A.8) and (A.11) that inequality (A.2) holds.

Case 4: $a_1 \leq a_7 \leq a_3$.

The proof in this case is similar to that in Case 3 and so we are done.